\documentclass[journal]{IEEEtran}

\usepackage{graphicx}
\usepackage{amsmath}
\usepackage{algorithmic}

\usepackage{array}
\usepackage{url}
\usepackage{authblk}
\usepackage{hyperref}
\usepackage{booktabs}
\usepackage{multirow}

\begin{document}

\title{The Extreme Cardiac MRI Analysis Challenge under Respiratory Motion (CMRxMotion)}

\author[1,2*]{Shuo Wang\thanks{\\* Task 1 contact: \href{mailto:shuowang@fudan.edu.cn}{shuowang@fudan.edu.cn} \\ $\dag$ Task 2 contact: \href{mailto:c.qin15@imperial.ac.uk}{c.qin15@imperial.ac.uk}\\ $\ddagger$ CMR acquisition contact: \href{mailto:wangcy@fudan.edu.cn}{wangcy@fudan.edu.cn}}}
\author[3$\dag$]{Chen Qin}
\author[4$\ddagger$]{Chengyan Wang}
\author[1,2]{Kang Wang}
\author[1,2]{Haoran Wang}
\author[5]{Chen Chen}
\author[5]{Cheng Ouyang}
\author[5]{Xutong Kuang}
\author[5]{Chengliang Dai}
\author[6]{Yuanhan Mo}
\author[7]{Zhang Shi}
\author[7]{Chenchen Dai}
\author[1,2]{Xinrong Chen}
\author[8]{He Wang}
\author[4]{Wenjia Bai}
\affil[1]{Digitial Medical Research Center, Fudan Univeristy, China}
\affil[2]{Shanghai Key Laboratory of MICCAI, China}
\affil[3]{Department of Electrical and Electronic Engineering, Imperial College London, UK}
\affil[4]{Human Phenome Institute, Fudan University, China}
\affil[5]{Department of Computing, Imperial College London, UK}
\affil[6]{School of Engineering, University of Oxford, UK}
\affil[7]{Department of Radiology, Zhongshan Hospital Affiliated to Fudan University, China}
\affil[8]{Institute of Science and Technology for Brain-inspired Intelligence, Fudan University, China}

\IEEEspecialpapernotice{(Summary of CMRxMotion Challenge Design)}

\maketitle

\begin{abstract}
The quality of cardiac magnetic resonance (CMR) imaging is susceptible to respiratory motion artifacts. The model robustness of automated segmentation techniques in face of real-world respiratory motion artifacts is unclear. This manuscript describes the design of extreme cardiac MRI analysis challenge under respiratory motion (CMRxMotion Challenge). The challenge aims to establish a public benchmark dataset to assess the effects of respiratory motion on image quality and examine the robustness of segmentation models. The challenge recruited 40 healthy volunteers to perform different breath-hold behaviors during one imaging visit, obtaining paired cine imaging with artifacts. Radiologists assessed the image quality and annotated the level of respiratory motion artifacts. For those images with diagnostic quality, radiologists further segmented the left ventricle, left ventricle myocardium and right ventricle. The images of training set (20 volunteers) along with the annotations are released to the challenge participants, to develop an automated image quality assessment model (Task 1) and an automated segmentation model (Task 2). The images of validation set (5 volunteers) are released to the challenge participants but the annotations are withheld for online evaluation of submitted predictions. Both the images and annotations of the test set (15 volunteers) were withheld and only used for offline evaluation of submitted containerized dockers. The image quality assessment task is quantitatively evaluated by the Cohen’s kappa statistics and the segmentation task is evaluated by the Dice scores and Hausdorff distances.

\end{abstract}

\begin{IEEEkeywords}
Cardiac magnetic resonance, image quality assessment, image segmentation, respiratory motion artifacts, model robustness.
\end{IEEEkeywords}

\IEEEpeerreviewmaketitle

\begin{figure*}[t]
\centering
\includegraphics[width=14cm]{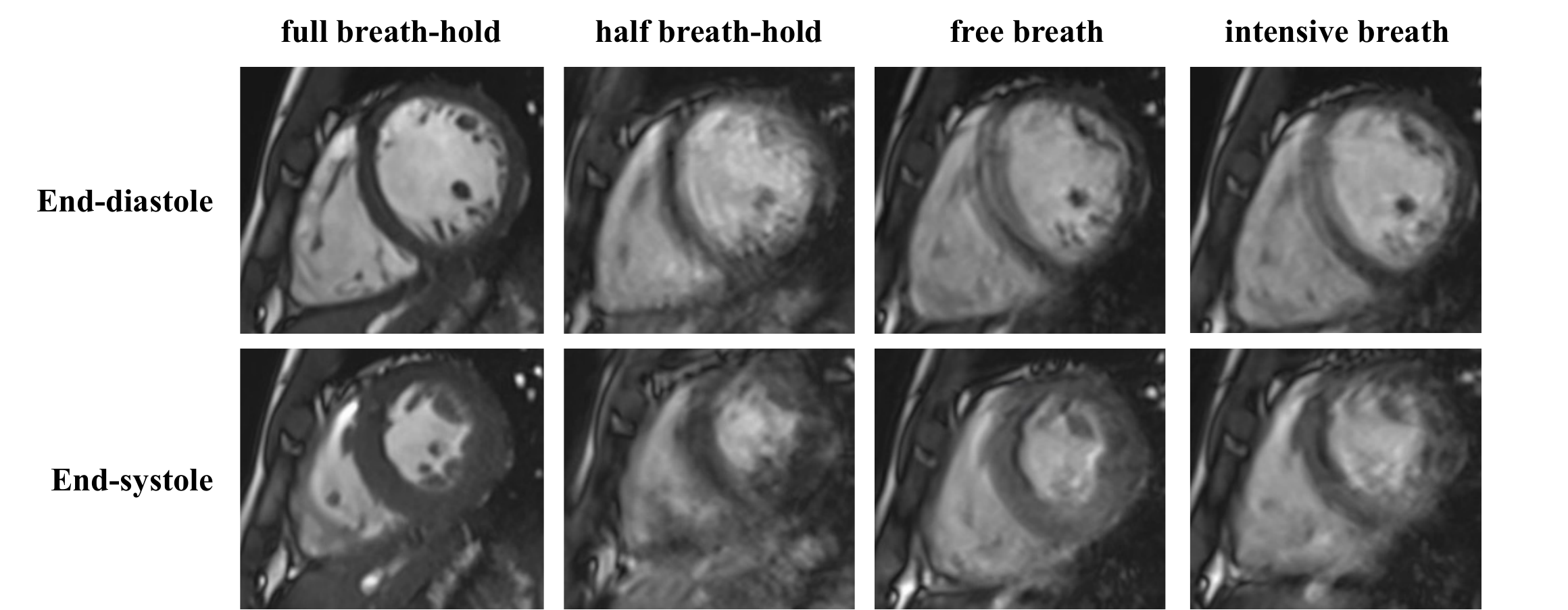}

\caption{Visual example of paired images acquired from the same volunteer performing four different breath-hold behaviors during CMR acquisition. End-diastole images are shown on the top row and end-systole images are shown on the bottom row.}
\label{fig1}
\end{figure*}

\section{Introduction}
Cardiac magnetic resonance (CMR) imaging is the current gold-standard modality for evaluating cardiac structure and function~\cite{schulz2020standardized}. Machine learning-based approaches have achieved remarkable performance in the CMR image segmentation task~\cite{chen2020deep,bai2020population}. However, the model performance is still challenged by inconsistent imaging environments (e.g., vendors and protocols)~\cite{campello2021multi}, population shifts (normal v.s. pathological cases)~\cite{bernard2018deep} and unexpected human behaviors (e.g., body movements) in clinical practice. It is useful to investigate potential failure modes~\cite{wang2020deep} by exposing a trained segmentation model to extreme cases in a ‘stress test’. To date, existing challenges focus on the vendor variability~\cite{campello2021multi} and anatomical structure variations~\cite{bernard2018deep} while the implications of human behaviors are less explored. For CMR image analysis, respiration motion is one of the major problems~\cite{ferreira2013cardiovascular}. Patients may not be able to follow the breath-hold instructions well, which is common in heart failure cases or children. The poor breath-hold behaviors result in degraded image quality and inaccurate analysis~\cite{wang2021joint}. Therefore, we launch the extreme cardiac MRI analysis challenge under respiratory motion (CMRxMotion Challenge) \footnote{\href{http://cmr.miccai.cloud}{http://cmr.miccai.cloud}}, to establish a public benchmark dataset to assess the effects of respiratory motion on CMR imaging quality and examine the robustness~\cite{paschali2018generalizability} of automated segmentation models. 

To curate such an extreme CMR dataset with respiratory motion artifacts, one retrospective way is to screen the images stored in hospital database and identify those problematic ones. But this requires considerable human efforts and may also bring about confounding factors such as vendors, scan protocol and pathology. Instead, we design a prospective study that health volunteers are recruited to perform different breath-hold behaviors during one imaging visit. As the confounding factors of MRI equipment and scan protocols are controlled, this extreme CMR dataset is established in specific to respiratory motion artifacts. The rest of the manuscript provides a summary of CMRxMotion challenge design, including the data acquisition, annotation, tasks, evaluation metric, ranking scheme and awards.

\section{Materials \& Methods}
In this challenge, we introduce the CMRxMotion dataset, a real-world cardiac MRI dataset including cases with extreme respiratory motions. We aim to provide such a CMR dataset that allows for a more comprehensive evaluation of model robustness under respiratory motion.

\subsection{Acquisition Design}
Forty volunteers are recruited for this study among the faculty and students of Fudan University. The volunteers are scanned on the same 3T MRI scanner (Siemens MAGNETOM Vida) at Zhangjiang International Brain Imaging Center, Fudan University. We follow the recommendations of CMR scans reported in the previous publication~\cite{wang2021recommendation}. The clinical CMR sequence ‘TrueFISP’ was used for bssfp cine imaging. For this challenge, we provide the short-axis (SA) images at the end-diastole (ED) and end-systole (ES) frames. Typical scan parameters are: spatial resolution 2.0×2.0 mm$^2$, slice thickness 8.0 mm, and slice gap 4.0 mm. The standard of procedure (SOP) requires volunteers to follow breath-hold instructions.

To induce respiratory motion artifacts, we attack the SOP by re-designing the participant’s breath-hold behaviors during CMR acquisition. Volunteers are trained to perform four different breath-hold behaviors, respectively: a) adhere to the breath-hold instructions; b) halve the breath-hold period; c) breathe freely; and d) breathe intensively. The scan is repeated for four times under above settings during  a single imaging visit. Therefore, we obtain a set of paired CMR images under different levels of respiratory motion artifacts. As two 3D short-axis image volumes at ED and ES frames were extracted, each subject contributes eight image volumes. An example of motion-degraded CMR image is demonstrated in Figure~\ref{fig1}. 

\subsection{Pre-processing}
The short-axis cine images are anonymized and exported to NIFTI format from the DICOM files. 3D image volumes at ED and ES frames are extracted for the challenge. The orientation of image volumes is discarded for better compatibility with NIFTI viewers (e.g., ITK-SNAP). 

\subsection{Challenge Dataset}
The images are split into training (20 volunteers, 160 image volumes), validation (5 volunteers, 40 image volumes) and test (15 volunteers, 120 image volumes) sets on subject level. Annotations were manually generated by radiologists for the challenge tasks. The training set along with the annotations are released to the challenge participants for model development. The images of validation set are released to the participants but the annotations are withheld for online evaluation of submitted predictions. Both the images and annotations of the test set are withheld and only used for offline evaluation of submitted containerized dockers.

\section{Challenge Tasks}
Based on the curated CMR dataset, we proposed two tasks: to develop an automated image quality assessment model (Task 1) and an automated
segmentation model (Task 2).

\begin{table*}[!ht]
\centering
\caption{Assessment criteria for subjective image quality and label definition in Task 1.}
\label{table1}
\begin{tabular}{@{}cll@{}}
\toprule
\textbf{  Label }     & \textbf{Image quality}                     & \textbf{Details}                                                                                 \\ \midrule
\multirow{2}{*}{\textbf{1}} & Excellent                          & No   artifacts present                                                                           \\ \cmidrule(l){2-3} 
                            & More   than adequate for diagnosis & Minor artifacts present   but image quality somewhat reduced                                     \\ \midrule
\textbf{2}                  & Adequate   for diagnosis           & Minor   artifacts present and image quality somewhat reduced but still sufﬁcient for   diagnosis \\ \midrule
\multirow{2}{*}{\textbf{3}} & Questionable   for diagnosis       & Image   quality impaired by artifacts so diagnostic value of images is questionable              \\ \cmidrule(l){2-3} 
                            & Non-diagnostic                     & Image quality heavily   impaired by artifacts and readers not able to assess                     \\ \bottomrule
\end{tabular}
\end{table*}

\subsection{Task 1: CMR image quality assessment}

\subsubsection{Motivation} 
Images with severe respiratory motion artifacts are not eligible for diagnostics and should be re-acquired if possible. It is useful to develop an automated image quality assessment model to recognize images with bad quality. In this task, we expect the challenge participants to develop an image quality assessment model to evaluate respiratory motion artifacts.

\subsubsection{Annotation}
All image volumes are viewed in 3D Slicer\footnote{\href{http://www.slicer.org}{http://www.slicer.org}} and the image quality is scored by radiologists. Two radiologists (Z.S. \& C.D.) discuss and reach a consensus on the annotation protocol. The standard 5-point Likert scale is used as follows: excellent diagnostic quality (5), more than adequate for diagnosis (4), adequate for diagnosis (3), questionable for diagnosis (2), and non-diagnostic (1). For better reproducibility, three levels of motion artifacts are defined based on the original 5-point scores. Images with quality scores 4-5 are labeled as mild motion artifacts, images with quality score 3 are labeled as intermediate motion artifacts, and images with quality score 1-2 are labeled as severe motion artifacts, as shown in Table~\ref{table1}. Therefore, the task is to predict the level of motion artifacts:
\begin{itemize}
\item Label 1: mild motion
\item Label 2: intermediate motion
\item Label 3: severe motion
\end{itemize}

\subsubsection{Metrics}
Cohen’s kappa statistics is used to evaluate the model performance in this task. It is a common metric to measure the level of agreement between two raters or judges who each classify items into mutually exclusive categories~\cite{benchoufi2020interobserver}:

\begin{equation}
        \kappa = \frac{p_0-p_e}{1-p_e}
        \label{kappa}
    \end{equation}
    
where $p_0$ is the overall accuracy of the model and $p_e$ is the agreement between the predictions and the ground truth as if happening by chance. Participating teams should submit a containerized docker predicting a label  from {1, 2, 3}, corresponding to mild/intermediate/severe motion artifacts. We calculate the Cohen’s kappa between the submission and the manual annotations, with implementations in sklearn\footnote{\href{https://scikit-learn.org/stable/modules/generated/sklearn.metrics.cohen_kappa_score.html}{https://scikit-learn.org/}}. 

\subsubsection{Ranking scheme}
The final ranking on test set is generated by sorting the Cohen’s kappa. To assess whether the performance difference is significant, we generate bootstrap samples and use paired Wilcoxon test.

\subsection{Task 2: Robust CMR image segmentation}
\subsubsection{Motivation} 
Learning-based image segmentation models are prone to fail in face of unseen extreme images. In this task, we prepare an extreme dataset reflecting different levels of image degradation due to respiratory motion in clinical practice. We expect the challenge participants to develop a segmentation model robust to the respiratory motion artifacts.

\subsubsection{Annotation}
\begin{figure}[t]
\centering
\includegraphics[width=8cm]{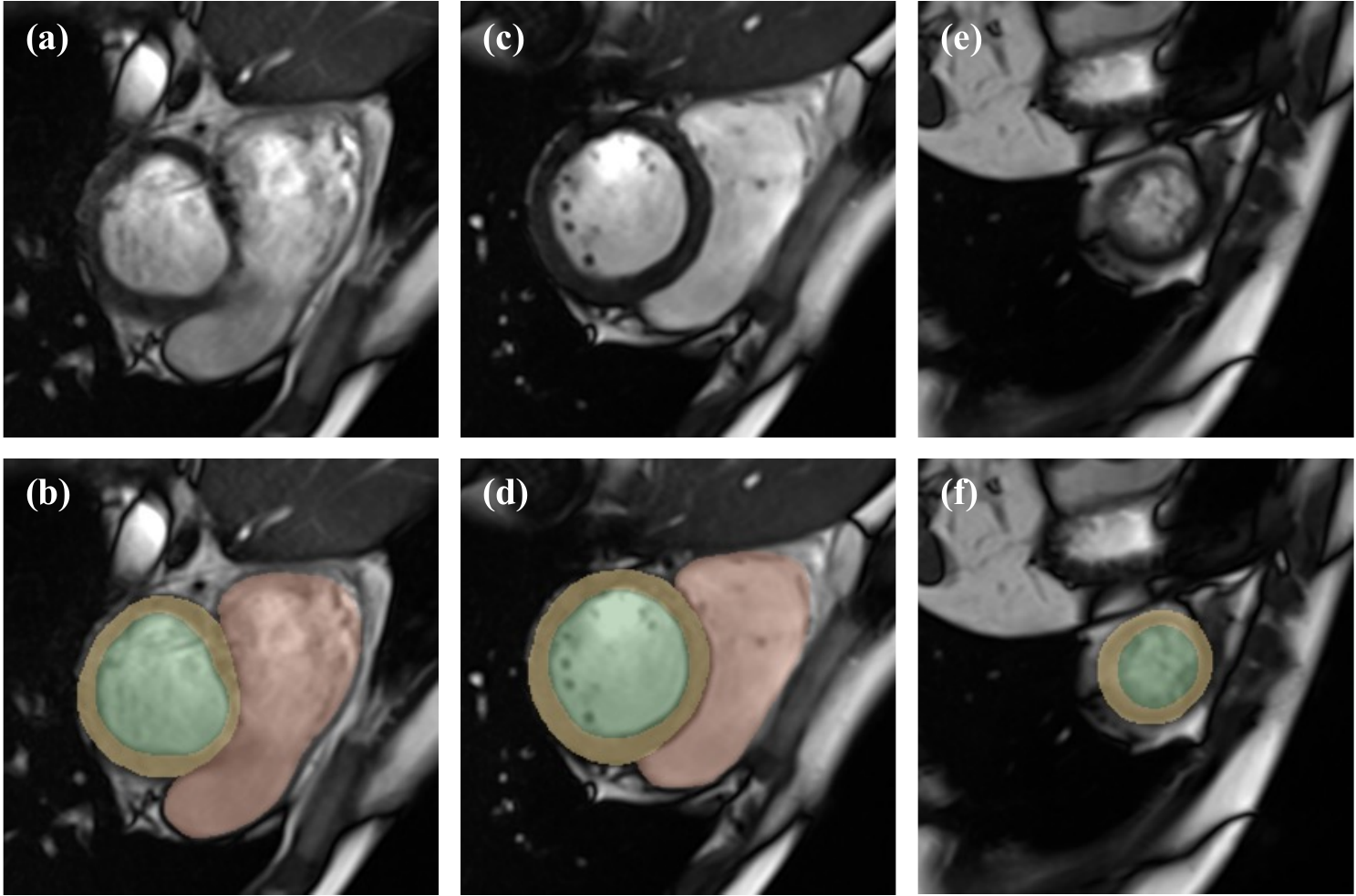}

\caption{Visual example of segmentation masks in Task 2. Basal slice image (a) and segmentation (b), mid-ventricle slice image (c) and segmentation (d), apical slice image (e) and segmentation (f) are shown. Left ventricle is in green, left ventricular myocardium is in yellow, and right ventricle is in red.}
\label{fig2}
\end{figure}

This challenge follows the annotation practice in previous CMR segmentation challenges~\cite{bernard2018deep,campello2021multi}, including contours for the left (LV) and right ventricle (RV) blood pools, as well as for the left ventricular myocardium (MYO). All images with diagnostic quality (i.e., Label 1 \& 2 in Task 1) in training, validation and test set are segmented by an experienced technician (X.K.) in 3D Slicer and reviewed by two radiologists (Z.S. and C.D.). Labels are: 0 (Background), 1 (LV), 2 (MYO) and 3 (RV). An example of segmentation task is shown in Figure~\ref{fig2}.

\subsubsection{Metrics}
Dice score and 95\% Hausdorff distance (HD95) are used to evaluate the model performance in this task. The Dice score measures the spatial overlap and is defined as

\begin{equation}
        Dice = \frac{2|X \cap Y|}{|X|+|Y|}
        \label{Dice}
    \end{equation}

where $X$ and $Y$ are the ground-truth and predicted segmentation masks, respectively.

The HD95 measures the contour difference between two segmentation masks, and is defined as

\begin{equation}
        HD95 = max\{\mathop{P_{95\%}}_{x\in X}d(x, Y), \mathop{P_{95\%}}_{y\in Y}d(y, X)\}
        \label{HD95}
    \end{equation}

where $d(x,Y)=\mathop{min}_{y\in Y} {\parallel}x-y{\parallel}$. Both Dice and HD95 metrics are calculated with implementations in MedPy\footnote{\href{http://loli.github.io/medpy/}{http://loli.github.io/medpy/}}.

\subsubsection{Ranking scheme}
Both the Dice scores and Hausdorff distances are used to compute the rankings. We take the ‘rank then aggregate’ strategy for the final ranking. For each case, we will compute the three Dice scores and three 95\% Hausdorff distance measures between the ground truth and the submitted segmentation of LV, MYO, and RV, respectively. For X number of cases included in the test set, each participant has X*6 rankings. The final ranking score is the average of all these rankings. The Wilcoxon test is used to check whether the ranking difference is significant.

\section{Participation and Awards}
\subsection{Registration}
Participants should firstly register on the challenge website\footnote{\href{http://cmr.miccai.cloud/register/}{http://cmr.miccai.cloud/register/}} and fill out the data access agreement. The imaging data and annotations can be downloaded on Synapse\footnote{https://www.synapse.org/\#!Synapse:syn28503327} by eligible teams.

\subsection{Submission}
For the training and validation phase, participating teams can submit their model prediction results for online evaluation on Synapse platform. For the test phase, participating teams should send their containerized dockers to the organizers for offline evaluation. The model performance on the test set is used for the final ranking.

\subsection{External Data}
Publicly available data is allowed while the source of external data must be provided.

\subsection{Awards}
Top three winners (including tier places) of each task receive monetary awards: 1st place \$500, 2nd place \$200, 3rd place \$100.

\section*{Acknowledgment}
We would like to thank all volunteers who kindly took time to participate in this study. We also thank Synapse and Paratera Technology Ltd for their cloud computing support.  

\newpage

\bibliographystyle{IEEEtran}
\bibliography{ref.bbl}

\end{document}